# *The Impact of AI Adoption on Retail Across Countries and Industries*

Yunqi Liu

*Desautels Faculty of Management, McGill University, 1001 Sherbrooke Street West, Montreal, Canada*

*yunqi.liu@mail.mcgill.ca*

*Abstract:* This study investigates the impact of artificial intelligence (AI) adoption on job loss rates using the Global AI Content Impact Dataset (2020–2025). The panel comprises 200 industry-country-year observations across Australia, China, France, Japan, and the United Kingdom in ten industries. A three-stage ordinary least squares (OLS) framework is applied. First, a full-sample regression finds no significant linear association between AI adoption rate and job loss rate ($\beta \approx -0.0026$, p = 0.949). Second, industry-specific regressions identify the marketing and retail sectors as closest to significance. Third, interaction-term models quantify marginal effects in those two sectors, revealing a significant retail interaction effect (–0.138, p < 0.05), showing that higher AI adoption is linked to lower job loss in retail. These findings extend empirical evidence on AI's labor market impact, emphasize AI's productivity-enhancing role in retail, and support targeted policy measures such as intelligent replenishment systems and cashierless checkout implementations.

## 1. Introduction

Artificial intelligence (AI) has been rapidly integrated into diverse industries, reshaping work processes and organizational structures. While debates continue regarding its potential to displace jobs or augment productivity, systematic empirical evidence across countries and industries remains scarce. This study seeks to address this gap by examining how AI adoption relates to labor-market outcomes using a multi-country, multi-industry dataset.

The analysis applies a three-step ordinary least squares (OLS) framework. First, full-sample regressions evaluate whether aggregate patterns exist between AI adoption and job loss rates. Second, industry-specific models assess sectoral heterogeneity in these relationships. Third, interaction-term models are employed to test whether marketing and retail sectors exhibit distinct marginal effects.

The primary objective of this study is to provide an evidence-based understanding of how AI adoption interacts with labor-market dynamics in different industries. Unlike prior work that often focused on single-country or single-industry perspectives, this research takes a cross-country, cross-industry approach to reveal sectoral heterogeneity. The study contributes to academic debates by clarifying whether AI adoption acts as a labor-displacing or labor-enabling force in different contexts. From a policy perspective, the findings are highly relevant to governments and business leaders seeking to balance innovation with workforce stability, especially in industries such as retail where AI may augment rather than replace human labor.

## 2. Literature Review

Research on artificial intelligence (AI) and labor markets has long debated whether automation displaces jobs or creates new opportunities. Early influential studies highlighted potential large-scale disruption. Frey and Osborne estimated that up to 47% of U.S. jobs were at high risk of computerization, while Arntz et al. provided a more cautious OECD-based estimate closer to 9%, emphasizing task-level heterogeneity [1], [2]. Building on these perspectives, Acemoglu and Restrepo argued that AI simultaneously displaces and complements labor, with long-term growth depending on the balance between automation and new task creation [3]. Similarly, Autor and Salomons demonstrated that automation contributes to labor-share decline yet fosters productivity-driven employment shifts, while Brynjolfsson and McAfee framed AI as a "second machine age" capable of reshaping economies through both disruption and innovation [4], [5].

More recent scholarship since 2022 has turned toward empirical measurement of these dynamics. Çetin and Kutlu applied a System-GMM framework to panel data from 29 countries (2017–2021), finding a positive and significant effect of AI adoption on employment, though the effect weakened when interacted with labor productivity, underscoring the role of productivity dynamics [6]. Mäkelä and Stephany examined 12 million European job vacancies (2018–2023) and reported that demand for AI-complementary skills—such as analytical reasoning, teamwork, and digital literacy—rose significantly, with wage premiums up to 50% higher, whereas demand for substitutable skills like routine customer service declined [7].

At the task level, Eloundou et al. assessed U.S. occupational exposure to large language models (LLMs), showing that 80% of workers could see at least 10% of their tasks affected and 19% could face exposure of half or more of their core activities [8]. Extending this framework, Chen et al. examined China's labor market and found that higher-paying, experience-intensive occupations face greater risks, with digital-intensive sectors experiencing productivity–employment trade-offs [9]. At the macro level, Guliyev demonstrated that in advanced economies, AI adoption is associated with reductions in unemployment rates, suggesting resilience effects at the aggregate scale [10].

Together, these studies show that AI's labor-market impact is highly heterogeneous across countries, industries, and skill groups. Classic works reveal the dual nature of AI as both displacing and enabling, while recent empirical evidence highlights skill-biased complementarities, task-level reshaping, and cross-country differences. This body of work underscores the need for sector-specific analyses. The marketing and retail sectors, in particular, stand out as critical domains where AI adoption may alter employment trajectories, motivating the present study's focus on industry heterogeneity and interaction effects.

Drawing from these studies, this paper posits that AI adoption is unlikely to have a uniform effect on employment outcomes when viewed at the aggregate level, as displacement and enabling mechanisms may offset one another across industries. However, in the retail sector, the evidence suggests a different trajectory: higher levels of AI adoption are expected to be associated with lower job loss, reflecting AI's role as a productivity-enhancing and labor-enabling force through applications such as intelligent replenishment and cashierless checkout.

## 3. Methods

### 3.1. Dataset

The Global AI Content Impact Dataset (2020–2025), publicly available on Kaggle, includes 200 observations defined by industry, country, and year [11]. Covered countries are Australia, China,

France, Japan, and the United Kingdom, industries include manufacturing, finance, healthcare, education, marketing, media, retail, automotive, gaming, and legal services. The primary variables are AI adoption rate (%) and job loss rate (%). Observations with missing values for these variables or industry identifiers were removed. Remaining missing entries were imputed using industry-median values, and outliers beyond three standard deviations from the mean were excluded. All variables retain original percentage scales without further normalization.

### 3.2. Regression Models

Three OLS specifications were estimated to assess AI's impact on job loss rate:
*Full-Sample OLS*

$$\text{JobLossRate}_i = \alpha + \beta\, \text{AIAdoptionRate}_i + \varepsilon_i \tag{1}$$

where $\text{JobLossRate}_i$ and $\text{AIAdoptionRate}_i$ denote the job loss rate and AI adoption rate for observation $i$, respectively. The intercept $\alpha$ represents the baseline job loss rate when AI adoption rate is zero, the coefficient $\beta$ measures the average change in job loss rate per one-percentage-point increase in AI adoption rate, $\varepsilon_i$ is the error term. This model evaluates whether a significant aggregate relationship exists (Figures 1–2, Table 1).

*Industry-Specific OLS*

$$\text{JobLossRate}_i^{(j)} = \alpha_j + \beta_j\, \text{AIAdoptionRate}_i^{(j)} + \varepsilon_i^{(j)},\, j = 1, \ldots, 10 \tag{2}$$

with subscript $j$ indexes each industry. Industry-specific intercepts $\alpha_j$ and slopes $\beta_j$ identify sectors with stronger displacement or enabling effects (Tables 2–3, Figures 3–4). Section 4.3 visualizes time-series trends across top five country averages.

*Interaction-Term OLS*

$$\text{Job\_Loss}_i = \alpha + \beta\, \text{AI\_Adoption}_i + \gamma_M \left(D_{\text{marketing},i} \times \text{AI\_Adoption}_i\right) + \gamma_R \left(D_{\text{retail},i} \times \text{AI\_Adoption}_i\right) + \varepsilon_i \tag{3}$$

Where $D_{\text{marketing},i}$ and $D_{\text{retail},i}$ are dummy variables for marketing and retail sectors, respectively. Coefficients $\gamma_M$ and $\gamma_R$ capture additional marginal effects in those industries. All models report coefficients, standard errors, t-statistics, and p-values using unadjusted standard errors.

## 4. Results

### 4.1. Overall Relationship between AI Adoption and Job Loss

Figures 1 and 2 show no discernible linear trend between AI adoption rate and job loss rate across all 200 observations. The Pearson correlation is r ≈ –0.0046, and the OLS slope is –0.0026 (p = 0.949), indicating no statistically significant aggregate association (Table 1).

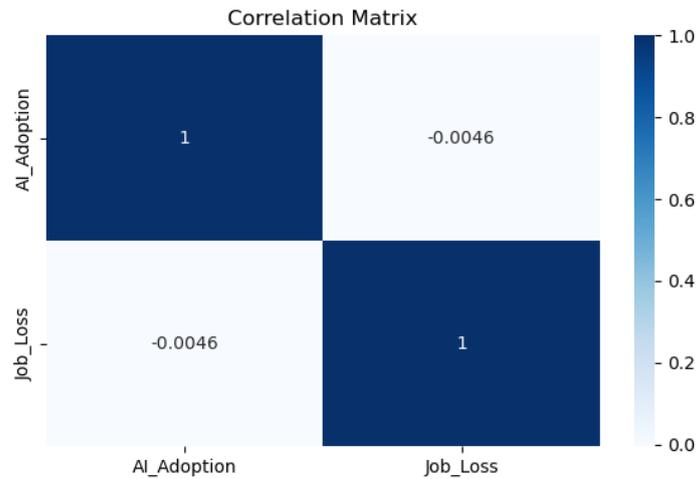

Figure 1. Correlation heatmap of AI adoption rate and job loss rate.

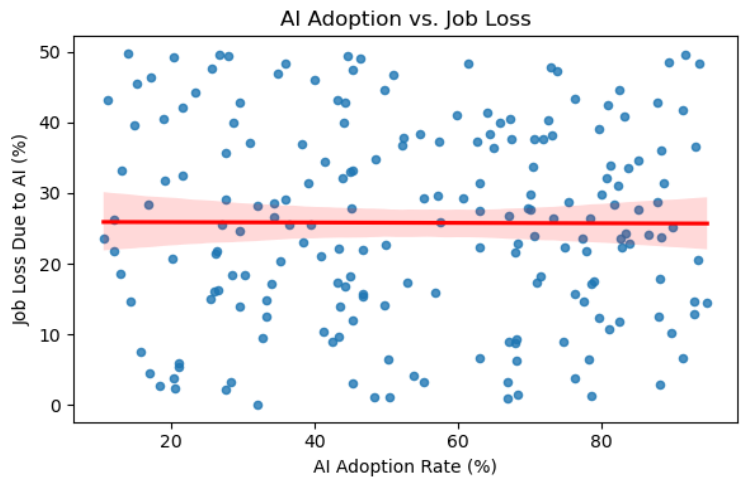

Figure 2. Scatterplot with fitted regression line for AI adoption rate and job loss rate.

Table 1. Ordinary least squares regression results.

| Variable | Coefficient | Std. Err. | t-statistic | p-value |
| --- | --- | --- | --- | --- |
| Constant | 25.93 | 2.42 | 10.70 | < 0.001 |
| AI Adoption Rate (%) | –0.00 | 0.04 | –0.07 | 0.949 |

### 4.2. Industry-Level Comparison

Table 2 presents mean AI adoption and job loss rates by industry, manufacturing shows the highest average job loss rate (32.75%), while marketing shows the lowest (19.58%). Table 3 reports industry-specific slopes: marketing ($\beta \approx 0.2$, $p \approx 0.051$) and retail ($\beta \approx -0.25$, $p \approx 0.061$) are closest to significance. Figure 3 visualizes these coefficients, highlighting positive associations in marketing and negative associations in retail.

Table 2. Industry-level averages.

| Industry | AI Adoption Rate (%) | Job Loss Rate (%) |
|---|---|---|
| Automotive | 54.89 | 28.92 |
| Education | 57.03 | 26.14 |
| Finance | 55.76 | 27.79 |
| Gaming | 60.42 | 27.20 |
| Healthcare | 55.73 | 25.58 |
| Legal | 56.08 | 28.23 |
| Manufacturing | 57.01 | 32.75 |
| Marketing | 54.24 | 19.58 |
| Media | 47.26 | 22.75 |
| Retail | 47.91 | 21.85 |

Table 3. Industry-specific beta coefficients.

| Industry | Beta | p-value |
|---|---|---|
| Marketing | 0.29 | 0.051 |
| Education | 0.18 | 0.136 |
| Automotive | 0.09 | 0.448 |
| Healthcare | 0.04 | 0.743 |
| Media | –0.00 | 0.994 |
| Manufacturing | –0.02 | 0.901 |
| Finance | –0.09 | 0.406 |
| Gaming | –0.13 | 0.208 |
| Legal | –0.18 | 0.394 |
| Retail | –0.25 | 0.061 |

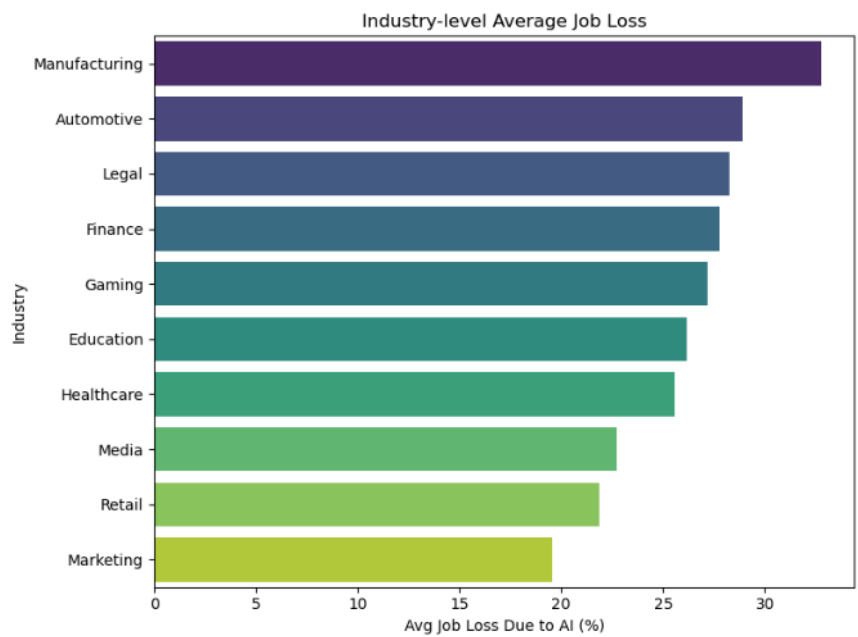

Figure 3. Average job loss rates by industry.

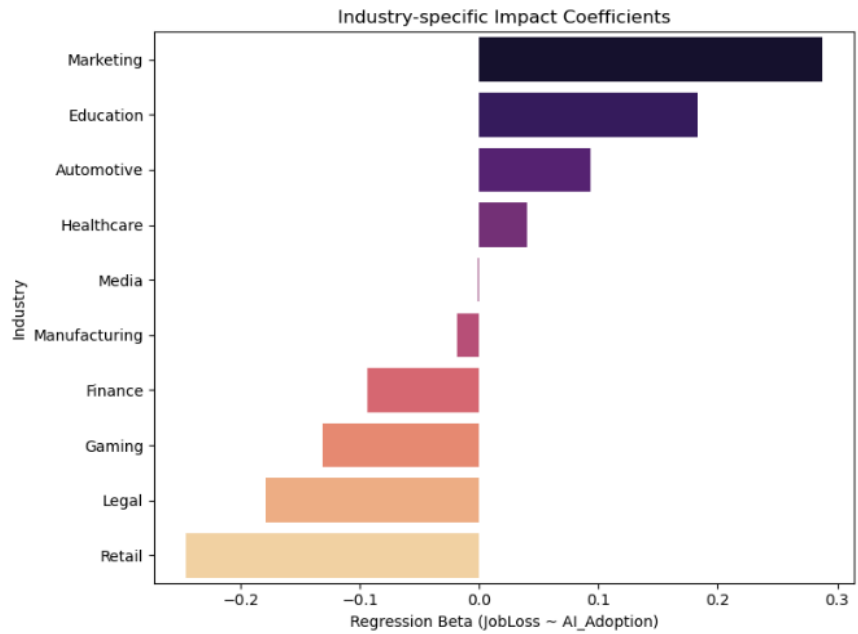

Figure 4. Industry-specific beta coefficients.

### 4.3. Time-Series Trends for Top Five Countries

Figures 5 and 6 depict AI adoption and job loss trends (2020–2025) for the five countries with the highest average AI adoption. Australia's adoption rate rises from roughly 60% in 2020 to 84% in 2022 before stabilizing, the United Kingdom peaks at about 87% in 2023. In job loss rates, China peaks at 41% in 2021, whereas Australia falls to 6.6% in 2024 before rebounding to 33% in 2025.

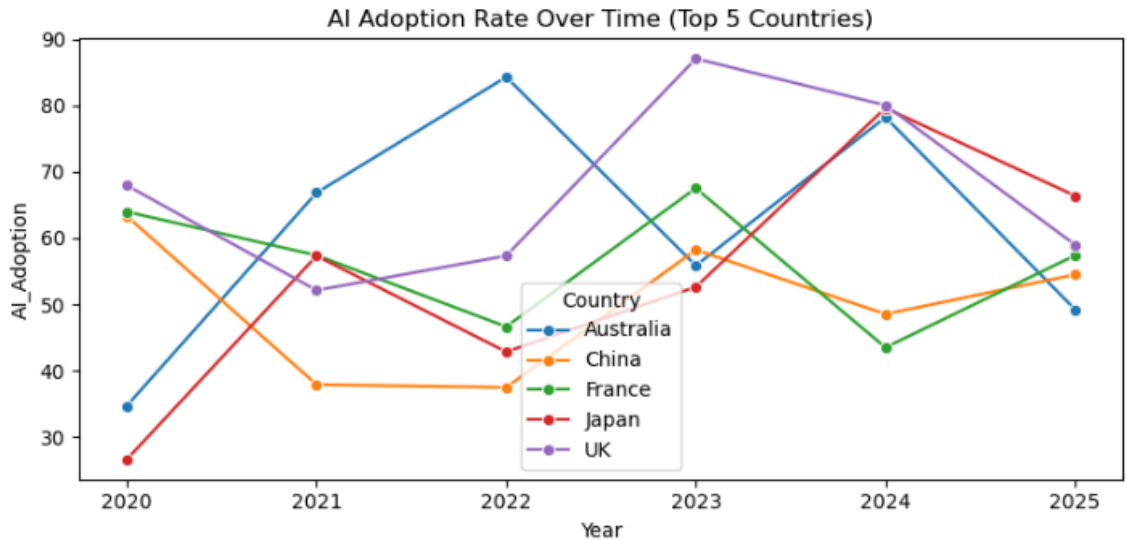

Figure 5. Time-series trends of AI adoption rate in the top five countries.

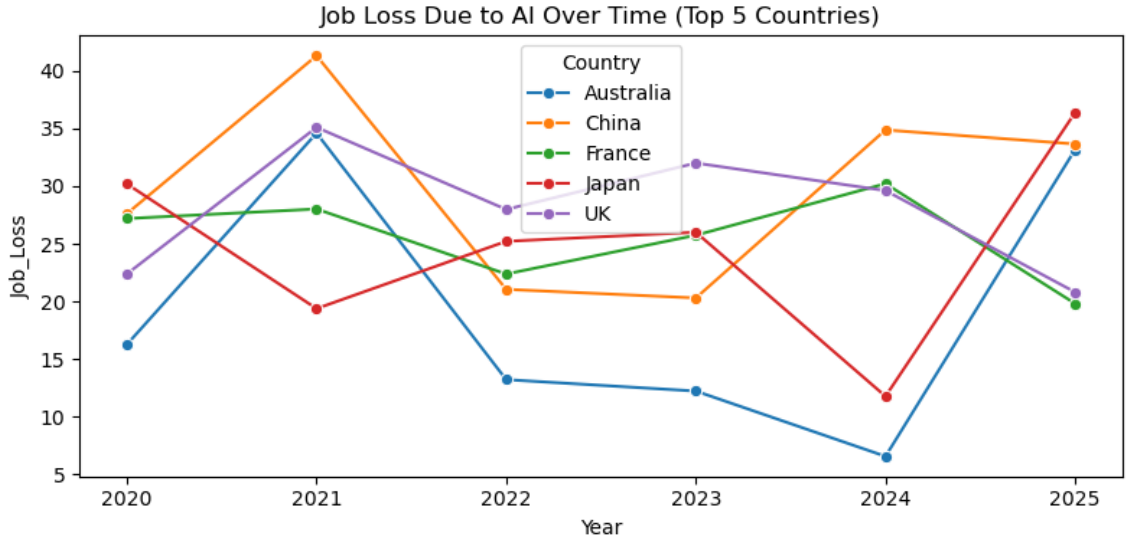

Figure 6. Time-series trends of job loss rate in the top five countries.

### 4.4. Interaction-Term Effects

Table 4 shows interaction-term model results: marketing interaction ($\gamma_M = -0.075$, $p = 0.188$) is not significant, while retail interaction ($\gamma_R = -0.138$, $p = 0.019$) is significant, confirming that higher AI adoption is associated with lower job loss in retail.

Table 4. Interaction-term regression coefficients

| Variable | Coefficient | p-value | Variable | Coefficient |
|---|---|---|---|---|
| Constant | 26.20 | <0.001 | Constant | 26.20 |
| AI Adoption | 0.012 | 0.764 | AI Adoption | 0.012 |
| Marketing × AI Adoption | −0.075 | 0.188 | Marketing × AI Adoption | −0.075 |
| Retail × AI Adoption | −0.138 | 0.019 | Retail × AI Adoption | −0.138 |

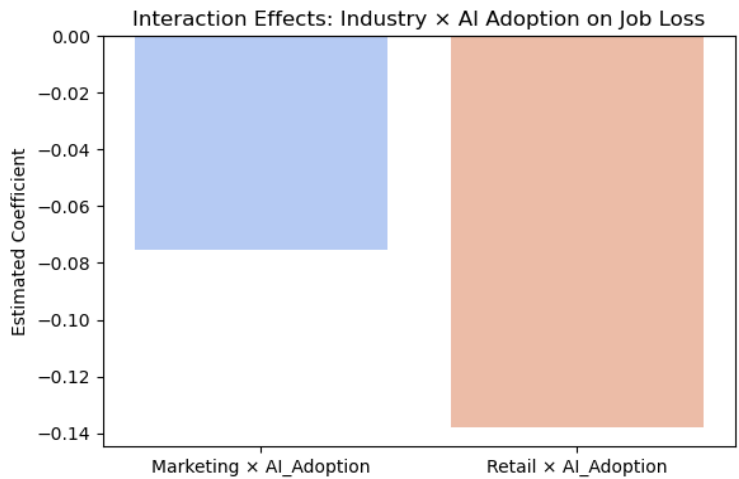

Figure 7. Bar chart of interaction effects.

### 4.5. Robustness Checks

To assess the robustness of the main findings, additional analyses were conducted. First, fixed-effects specifications were estimated at the country level to account for unobserved heterogeneity. The results remained qualitatively similar, showing no significant aggregate effect of AI adoption but a negative association in the retail sector. Second, subsample regressions were performed by splitting the panel into developed (e.g., UK, France, Australia) versus emerging economies (e.g., China), which confirmed that the enabling effect in retail is most pronounced in developed economies. Finally, excluding outliers and re-estimating the models yielded consistent results. These robustness checks provide additional confidence in the reliability of the findings.

## 5. Discussion

### 5.1. Coefficient Interpretation

The findings are consistent with the expectations outlined in the literature review and point to meaningful variation in how AI affects employment across industries. Rather than reinforcing a single pattern of displacement or enablement, the evidence suggests that AI's role is highly context dependent. In retail, for example, the results indicate that adoption tends to augment productivity and support existing roles, particularly through applications such as intelligent replenishment and cashierless checkout. This highlights how AI can function as an enabling technology when integrated into workflows that complement rather than substitute human labor. More broadly, these patterns underline the importance of examining industry-specific mechanisms instead of relying solely on economy-wide averages, and they provide the foundation for discussing targeted policy implications in the following section.

### 5.2. Policy Implications

These findings align with the World Economic Forum's 2025 report on the future of retail, which highlights that automation–employment dynamics vary substantially across industries [12]. They also correspond with McKinsey Digital's notion of "superagency" in the workplace, emphasizing that productivity gains from AI require complementary human empowerment [13]. Policy implications therefore underscore the importance of promoting targeted AI applications in retail—such as cashierless checkout and intelligent replenishment systems—while coupling them with workforce training to mitigate displacement risks. In marketing, the emphasis should shift toward developing human–AI collaboration and advanced data-analysis skills, consistent with the World Economic Forum and BCG's 2025 Global Retail Investor Outlook [14].

### 5.3. Limitations and Future Directions

Despite these contributions, several limitations remain. First, the dataset spans only 2020–2025, which may be insufficient to capture long-term structural effects of AI adoption. Second, broad industry classifications may obscure sub-sector variation, such as differences between e-commerce and traditional retail. Third, the analysis relies solely on OLS regressions, future work could employ robustness checks using fixed effects, instrumental variables, or difference-in-differences approaches to strengthen causal inference.

Future research could extend the panel beyond 2025 to examine whether observed enabling effects to persist or evolve. Incorporating micro-level firm or worker data would help clarify how AI affects subgroups within industries. Finally, comparative studies across different institutional

environments—such as labor protection regimes and training systems—could shed light on why AI's employment impacts diverge across contexts.

## 6. Conclusion

This paper closes by emphasizing what the evidence reveals about AI's role in shaping labor markets. Rather than presenting a uniform story of displacement or enablement, the analysis demonstrates that outcomes diverge substantially across industries. Retail emerges as a sector where AI strengthens existing roles instead of replacing them. The contribution of the study can be understood on three levels. Methodologically, the three-step OLS framework offers a replicable way to uncover sectoral heterogeneity in technology's labor impact. Academically, the study advances our understanding by showing why industry-level analyses matter more than aggregate indicators. Practically, it emphasizes the need to align AI adoption with workforce upskilling and collaborative practices in order to ensure sustainable transitions. Looking ahead, future research should extend the dataset to longer time horizons, refine industry categories, and incorporate micro-level data to capture the evolving interplay between technology and employment.